# Unitary Differential Space-Time Modulation with Joint Modulation

C. Yuen, Y. L. Guan, T. T. Tjhung

**Abstract** – We develop two new designs of unitary differential space-time modulation (DSTM) with low decoding complexity. Their decoder can be separated into a few parallel decoders, each of which has a decoding search space of less than $\sqrt{N}$ if the DSTM codebook contains $N$ codewords. Both designs are based on the concept of joint modulation, which means that several information symbols are jointly modulated, unlike the conventional symbol-by-symbol modulation. The first design is based on Orthogonal Space-Time Block Code (O-STBC) with joint constellation constructed from spherical codes. The second design is based on Quasi-Orthogonal Space-Time Block Code (QO-STBC) with specially designed pair-wise constellation sets. Both the proposed unitary DSTM schemes have considerably lower decoding complexity than many prior DSTM schemes, including those based on Group Codes and Sp(2) which generally have a decoding search space of $N$ for a codebook size of $N$ codewords, and much better decoding performance than the existing O-STBC DSTM scheme. Between two designs, the proposed DSTM based on O-STBC generally has better decoding performance, while the proposed DSTM based on QO-STBC has lower decoding complexity when 8 transmit antennas.

## I. INTRODUCTION

Modulation techniques designed for multiple transmit antennas, called space-time modulation or transmit diversity can be used to reduce fading effects effectively. Early transmit diversity schemes were designed for coherent detection, with channel estimates assumed available at the receiver. However, the complexity and cost of channel estimation grow with the number of transmit and receive antennas. Therefore, transmit diversity schemes that do not require channel estimation are desirable. To this end, several differential space-time modulation (DSTM) schemes have been proposed [1-7]. The DSTM schemes in [1-6] generally have a decoding search space of $N$ for a DSTM codebook with $N$ codewords due to the lack of orthogonality or quasi-orthogonality in the code structure. This leads to an exponential increase in decoding complexity with spectral efficiency. For instance, in order to provide a spectral efficiency of 1.5bps/Hz, the codebook of the DSTM in [1,2] for four transmit antennas has $N$

= $2^6$ = 64 codewords, hence its optimal decoder needs to search over a space of 64. This is increased to $2^8$ = 256 if the spectral efficiency is increased to 2bps/Hz. On the other hand, the scheme in [7] is single-symbol decodable as it is designed based on the square Orthogonal Space-Time Block Code (O-STBC). Hence for the same four transmit antennas and spectral efficiency of 1.5bps/Hz, the decoding search space per decoder of O-STBC DSTM could be as low as 4, or $\sqrt[3]{N}$ where $N$ = 64 is the codebook size. However, such reduction in decoding complexity is obtained with a sacrifice in the decoding performance, and its maximum achievable code rate is limited to ¾ for four antennas and ½ for eight antennas. To trade decoding complexity for performance, a non-unitary DSTM scheme based on the Quasi-Orthogonal STBC (QO-STBC) and a unitary DSTM scheme based on unitary non-linear STBC have been proposed in [15] and [16] respectively. Both support full rate (code rate = 1 symbol/channel use) for four transmit antennas, and both are pair-wise decodable with a decoding search space in between $\sqrt{N}$ and $\sqrt[3]{N}$ per decoder.

In this paper, we propose two new unitary DSTM schemes based on the concept of *joint modulation*. Their decoding can be performed by two or three parallel decoders, with a search space of less than $\sqrt{N}$ per decoder. The first design is based on unitary matrices derived from O-STBC with joint constellation constructed from *spherical codes*. The second design is based on unitary matrices derived from double-symbol-decodable QO-STBC [8-13] with a pairwise constellation set.

Compared with the scheme from [15], which is also based on QO-STBC, our proposed schemes are unitary DSTM designs, while that in [15] is not. The DSTM encoder of [15] also has higher computational complexity as it needs to solve a set of linear equations for every codeword to be transmitted. Compared with the schemes from [16], our proposed designs can be extended to pairwise-decodable QO-STBCs of any number of transmit antennas; it is also able to support a wide range of spectral efficiencies; while those in [16] are not known to exhibit such flexiblity.

## II. REVIEW OF UNITARY DSTM

*A. Unitary DSTM Signal Model*

Consider a MIMO communication system with $N_T$ transmit and $N_R$ receive antennas. Let $\mathbf{H}_t$ be the $N_R \times N_T$ channel gain matrix at a time $t$. Let $\mathbf{C}_t$ be the $N_T \times P$ codeword transmitted at a time $t$. Then, the received signal matrix $\mathbf{R}_t$ can be written as

$$\mathbf{R}_t = \mathbf{H}_t \mathbf{C}_t + \mathbf{N}_t \tag{1}$$

where $\mathbf{N}_t$ is the additive white Gaussian noise. At the start of the transmission, we transmit a known codeword $\mathbf{C}_0$, which is a unitary matrix. The codeword $\mathbf{C}_t$ transmitted at a time $t$ is differentially encoded by

$$\mathbf{C}_t = \mathbf{C}_{t-1} \mathbf{U}_t \tag{2}$$

where $\mathbf{U}_t$ is a unitary matrix of size $N_T \times N_T$ (such that $\mathbf{U}_t \mathbf{U}_t^H = \mathbf{I}$), called the *code matrix*, that contains information of the transmitted data. If we assume that the channel remains unchanged during two consecutive code periods, i.e. $\mathbf{H}_t = \mathbf{H}_{t-1}$, the received signal $\mathbf{R}_t$ at a time $t$ can be expressed [7] as

$$\mathbf{R}_t = \mathbf{H}_t \mathbf{C}_{t-1} \mathbf{U}_t + \mathbf{N}_t = (\mathbf{R}_{t-1} - \mathbf{N}_{t-1}) \mathbf{U}_t + \mathbf{N}_t = \mathbf{R}_{t-1} \mathbf{U}_t + \tilde{\mathbf{N}}_t \tag{3}$$

where $\tilde{\mathbf{N}}_t = -\mathbf{N}_{t-1} \mathbf{U}_t + \mathbf{N}_t$ is an equivalent additive white Gaussian noise. The corresponding decision metric for (3) is,

$$\hat{\mathbf{U}}_t = \arg\min_{\mathbf{U}_t \in \mathcal{U}} \text{tr}\left(\{\mathbf{R}_t - \mathbf{R}_{t-1}\mathbf{U}_t\}^H \{\mathbf{R}_t - \mathbf{R}_{t-1}\mathbf{U}_t\}\right) = \arg\max_{\mathbf{U}_t \in \mathcal{U}} \text{Re}\left\{\text{tr}\left(\mathbf{R}_t^H \mathbf{R}_{t-1} \mathbf{U}_t\right)\right\} \tag{4}$$

where $\mathcal{U}$ denotes the set of all possible code matrices.

*B. Diversity and Coding Gain*

The design criteria of unitary DSTM scheme have been formulated in [1] and found to be the same as those for coherent space-time coding. The transmit diversity level that can be achieved is given by:

$$\text{Min}\left[\text{rank}(\mathbf{U}_k - \mathbf{U}_l)\right] \quad \forall k \neq l. \tag{5}$$

In order to achieve full transmit diversity, the minimum rank in (5) has to be equal to $N_T$ and the DSTM code is said to be of full rank. For a full-rank unitary DSTM code, its *coding gain* is defined in [1, 7] as

$$\text{Min}\left[N_T \times \det\left((\mathbf{U}_k - \mathbf{U}_l)(\mathbf{U}_k - \mathbf{U}_l)^H\right)^{1/N_T}\right] \quad \forall k \neq l. \tag{6}$$

In order to achieve optimum decoding performance, the coding gain has to be maximized.

### III. A NEW UNITARY DSTM SCHEME BASED ON O-STBC

In this section, we shall develop a new unitary DSTM scheme using the well-known square O-STBC. For simplicity, we will use the rate-3/4 O-STBC for four transmit antennas described in [17] as an example. The proposed unitary DSTM technique is applicable to any square O-STBC for any number of transmit antennas.

*A. Orthogonal Space-Time Block Code*

The 4×4 codeword of rate-3/4 O-STBC in [17] (herein denoted as $\mathbf{C_{O4}}$) is shown in (7):

$$\mathbf{C_{O4}} = \begin{bmatrix} c_1 & 0 & c_2 & -c_3 \\ 0 & c_1 & c_3^* & c_2^* \\ -c_2^* & -c_3 & c_1^* & 0 \\ c_3^* & -c_2 & 0 & c_1^* \end{bmatrix}; \quad \mathbf{C_{O4}}\mathbf{C_{O4}^H} = \begin{bmatrix} \alpha & 0 & 0 & 0 \\ 0 & \alpha & 0 & 0 \\ 0 & 0 & \alpha & 0 \\ 0 & 0 & 0 & \alpha \end{bmatrix} \tag{7}$$

where $c_i$, $1 \leq i \leq 3$, represents the complex information symbol to be transmitted and

$$\alpha = \sum_{i=1}^{3} |c_i|^2 \tag{8}$$

The determinant of the codeword distance matrix of $\mathbf{C_{O4}}$ can be easily shown to be:

$$\det = \left( \sum_{i=1}^{3} |\Delta_i|^2 \right)^4 \tag{9}$$

where $\Delta_i$, $1 \leq i \leq 3$, represents the possible error in the $i^{th}$ transmitted constellation symbol. O-STBC can always achieve full diversity as (9) can never become zero as long as $\Delta_i$ is not zero for all $i$.

*B. New Unitary DSTM Scheme Based on O-STBC with Joint Modulation*

$\mathbf{C_{O4}}$ in (7) can be used as the unitary code matrix of a unitary DSTM scheme as long as $\alpha$ is equal to 1. The DSTM approach taken in [7] is to select the symbols $c_i$ from a PSK constellation (that has a constant power) to conform to (7). In order to achieve a better performance, we let multiple symbols be *jointly modulated* in order to conform to (7). Specifically, we separate the data into two groups with three real symbols per group, and jointly modulate these three real symbols. This suggests that $c_1^R, c_1^I, c_2^R$ could be jointly mapped to a tri-symbol $\{a_k, b_k, c_k\}$, while $c_2^I, c_3^R, c_3^I$ are mapped to another tri-symbol {$a_l$, $b_l$, $c_l$}, such that $|a_k|^2 + |b_k|^2 + |c_k|^2 = |a_l|^2 + |b_l|^2 + |c_l|^2 = 0.5$ for all values of $k$ and $l$. To maximize the coding gain, the symbol-pairs should further be designed to maximize the minimum value of det in (9).

In short, in this scheme we use $\mathbf{C_{O4}}$ with code symbols drawn from a special joint constellation set $\mathcal{M}$ consisting of complex-valued symbol-pairs {$a_k$, $b_k$, $c_k$} that satisfy the following criteria:

(i)  Power Criterion: $|a_k|^2 + |b_k|^2 + |c_k|^2 = 0.5$     (10)

(ii) Performance Criterion: maximize $\text{Min}\{|\Delta a_{kl}|^2 + |\Delta b_{kl}|^2 + |\Delta c_{kl}|^2\}$

where $\Delta a_{kl} = a_k - a_l$, $\Delta b_{kl} = b_k - b_l$, and $\Delta c_{kl} = c_k - c_l$ for all $k \neq l$. The systematic design of $\mathcal{M}$ will be elaborated in the next section.

The spectral efficiency, *Eff*, of the resultant unitary DSTM scheme based on rate-3/4 O-STBC is *Eff* = $2(\log_2 L)/N_T$ bps/Hz where $L$ is the total number of symbol-pairs {$a_k$, $b_k$, $c_k$} in $\mathcal{M}$. For example, consider the case of four transmit antennas ($N_T = 4$) and a target spectral efficiency of *Eff* = 1.5 bps/Hz. From the *Eff* expression above, the required constellation size is $L = 8 = 2^3$. Therefore, in the *encoder* of the proposed DSTM scheme, a constellation set $\mathcal{M}$ with 8 tri-symbol's will first have to be designed according to (10). Every first three information bits will be mapped to a tri-symbol {$a_k$, $b_k$, $c_k$} drawn from $\mathcal{M}$ to constitute the code symbols {$c_1^R, c_1^I, c_2^R$} in $\mathbf{C_{O4}}$ in (7), while every next three information bits will be mapped to another symbol-pair {$a_l$, $b_l$, $c_l$} drawn from $\mathcal{M}$ to constitute the code symbols {$c_2^I, c_3^R, c_3^I$} in $\mathbf{C_{O4}}$.

In the *decoder* of the proposed DSTM scheme, with $\mathbf{U_t}$ in (4) set to $\mathbf{C_{O4}}$ in (7), the decision metrics in (4) can be simplified to:

$$\begin{aligned}\{\hat{c}_1^R, \hat{c}_1^I, \hat{c}_2^R\} &= \underset{\{c_1^R, c_1^I, c_2^R\} \in \mathcal{M}}{\arg\max} \left[ \sum_{i=1,2} \text{Re}\{\text{tr}(\mathbf{R}_t^H \mathbf{R}_{t-1} \mathbf{A}_i)\} c_i^R + \text{Re}\{\text{tr}(\mathbf{R}_t^H \mathbf{R}_{t-1} j\mathbf{B}_1)\} c_1^I \right] \\ \{\hat{c}_2^I, \hat{c}_3^R, \hat{c}_3^I\} &= \underset{\{c_2^I, c_3^R, c_3^I\} \in \mathcal{M}}{\arg\max} \left[ \text{Re}\{\text{tr}(\mathbf{R}_t^H \mathbf{R}_{t-1} \mathbf{A}_3)\} c_3^R + \sum_{i=2,3} \text{Re}\{\text{tr}(\mathbf{R}_t^H \mathbf{R}_{t-1} j\mathbf{B}_i)\} c_i^I \right] \end{aligned}$$ (11)

where $\mathbf{A}_k$ and $\mathbf{B}_k$ are the dispersion matrices of $\mathbf{C_{O4}}$ [17].

The above shows that the proposed DSTM scheme can be decoded by jointly detecting the three real symbols {$c_1^R, c_1^I, c_2^R$}, then the other three real symbols {$c_2^I, c_3^R, c_3^I$}. Hence the search space is the square root of those reported in [1- 6]. In general, this DSTM achieves a decoding search space of $\sqrt{N}$ per decoder for a codebook with *N* codewords.

*C. Design of Joint Constellation Set from Spherical Code*

To design the joint constellation set $\mathcal{M}$ of tri-symbol to meet the conditions in (10), we note from (10)(i) that the constellation points must lie on the surface of a 3-dimensional sphere, and from (10)(ii) that the constellation points must be spaced as far apart as possible (i.e. the minimum distance between them is maximized). This falls under the realm of *spherical code*.

*Spherical code* (or *spherical packing*) deals with the problem of distributing *n* points on a sphere in *d* dimension such that the minimum distance (or equivalently the minimal angle) between any pair of points is maximized, and the maximum distance is called the *covering radius*. For the above DSTM example of four transmit antennas with spectral efficiency of 1.5bps/Hz, we need a spherical code with 3 dimensions and eight points (i.e. eight sets of tri-symbol).

A list of optimal spherical code has been found in [18]. In Table 1 we list the dimension *d*, number of points *n*, and minimum angular separation $\theta_{min}$ of a few spherical codes. The exact configurations of the spherical code with dimension 3 and 16 points are shown in Appendix A for illustration.

Table 1 Minimal separation for some optimal spherical codes

| Dimension *d* | Points *n* | Minimum separation $\theta_{min}$ (degree) |
|---|---|---|
| 3 | 8 | 74.8585 |
| 3 | 16 | 52.2444 |
| 4 | 64 | 42.3062 |

### IV. A NEW UNITARY DSTM SCHEME BASED ON QO-STBC

The O-STBC DSTM scheme proposed in Section III does not have full rate due to the complex O-STBC used. We are interested to know how a DSTM scheme based on QO-STBC, which is known to support a higher code rate than O-STBC, will perform. Of course, for a fair comparison, both DSTM schemes need to have the same or similar decoding complexity, which in this paper means a decoding search space less than $\sqrt{N}$ per decoder for a codebook with $N$ codewords.

In this section, we propose another new unitary DSTM scheme based on rate-1 double-symbol-decodable QO-STBC. The proposed technique is applicable to any double-symbol-decodable square QO-STBC, such as those designed in [8-11]. Here we will use the QO-STBC in [8] for four transmit antennas as an example.

*A. Quasi-Orthogonal Space-Time Block Code*

The 4 × 4 codeword of the QO-STBC in [8] (herein denoted as $\mathbf{C_{Q4}}$) is shown in (12):

$$\mathbf{C_{Q4}} = \begin{bmatrix} c_1 & -c_2^* & -c_3^* & c_4 \\ c_2 & c_1^* & -c_4^* & -c_3 \\ c_3 & -c_4^* & c_1^* & -c_2 \\ c_4 & c_3^* & c_2^* & c_1 \end{bmatrix}; \quad \mathbf{C_{Q4}C_{Q4}^H} = \begin{bmatrix} \alpha & 0 & 0 & \beta \\ 0 & \alpha & -\beta & 0 \\ 0 & -\beta & \alpha & 0 \\ \beta & 0 & 0 & \alpha \end{bmatrix} \quad (12)$$

where $c_i$, $1 \leq i \leq 4$, represents the complex information symbol to be transmitted and

$$\alpha = \sum_{i=1}^{4} |c_i|^2; \quad \beta = 2 \times Re(c_1 \times c_4^* - c_2 \times c_3^*). \quad (13)$$

Following [9-13], the minimum determinant value of $\mathbf{C_{Q4}}$ is obtained by assuming half of the codeword errors to be zero, i.e.

$$\det_{min} = \left[ (|\Delta_1 + \Delta_4|^2) \times (|\Delta_1 - \Delta_4|^2) \right]^2 \qquad \text{assuming } \Delta_2 = \Delta_3 = 0 \quad (14)$$

In order to achieve full transmit diversity and optimum coding gain, the value of $\det_{min}$ in (14) has to be non-zero and maximized.

*B. New Unitary DSTM Scheme Based on QO-STBC with Joint Modulation*

To use the $\mathbf{C_{Q4}}$ in (12) as the unitary code matrix of a unitary DSTM scheme, its $\alpha$ and $\beta$ values in (13) must be equal to 1 and 0 respectively, i.e. $\mathbf{C_{Q4}}$ must be unitary. However, unlike O-STBC, generally $\beta = 0$ cannot be achieved in QO-STBC if $c_1$ to $c_4$ are conventional independent PSK or QAM symbols. To have $\beta = 0$, we can see from (13) that $Re(c_1 c_4^*)$ must be equal to $Re(c_2 c_3^*)$. This suggests that $c_1$ and $c_4$ should be jointly mapped to a symbol-pair $\{a_k, b_k\}$, while $c_2$ and $c_3$ should be mapped to another symbol-pair $\{a_l, b_l\}$, such that $Re(a_k b_k^*) = Re(a_l b_l^*)$ for all $k$ and $l$. To further achieve $\alpha = 1$,

Eq.(13) shows that these symbol-pairs must further satisfy $|a_k|^2 + |b_k|^2 + |a_l|^2 + |b_l|^2 = 1$, or $|a_k|^2 + |b_k|^2 = |a_l|^2 + |b_l|^2 = 0.5$ if all symbol-pairs are required to have equal power. In addition, to maximize the coding gain, the symbol-pairs should further be designed to maximize the value of $\det_{min}$ in (14).

Summarizing the above, we conclude that to make the QO-STBC codeword $\mathbf{C_{Q4}}$ unitary, its code symbols must be drawn from a special pairwise/joint constellation set $\mathcal{M}$ which consists of complex-valued symbol-pairs $\{a_k, b_k\}$ that satisfy the following criteria:

(i) Unitary Criterion: $\mathrm{Re}(a_k b_k^*) = v$

(ii) Power Criterion: $|a_k|^2 + |b_k|^2 = 0.5$ (15)

(iii) Performance Criterion: maximize $\mathrm{Min}\left\{\left[|\Delta a_{kl} + \Delta b_{kl}|^2 \times |\Delta a_{kl} - \Delta b_{kl}|^2\right]^2\right\}$

where $v$ can be any constant value, and $\Delta a_{kl} = a_k - a_l$, $\Delta b_{kl} = b_k - b_l$ for all $k \neq l$. The spectral efficiency, *Eff*, of the resultant unitary DSTM scheme based on full-rate QO-STBC is $Eff = 2(\log_2 L)/N_T$ bps/Hz where $L$ is the total number of symbol-pairs $\{a_k, b_k\}$ in $\mathcal{M}$.

For example, consider a system with four transmit antennas ($N_T = 4$) and a target spectral efficiency of $Eff = 2$ bps/Hz. From the equation of *Eff*, the required constellation size is $L = 16 = 2^4$. In the *encoder* of the proposed DSTM scheme, a pairwise constellation set $\mathcal{M}$ with 16 symbol-pairs will first have to be designed according to (15). Four information bits will be mapped to a symbol-pair $\{a_k, b_k\}$ in $\mathcal{M}$ to constitute the code symbols $\{c_1, c_4\}$ in $\mathbf{C_{Q4}}$ in (12), while another four information bits will be mapped to another symbol-pair $\{a_l, b_l\}$ in $\mathcal{M}$ to constitute the code symbols $\{c_2, c_3\}$ in $\mathbf{C_{Q4}}$. With such pairwise constellation design, the resultant $\mathbf{C_{Q4}}$ will be unitary and can now be used as the unitary DSTM code matrix $\mathbf{U}_t$.

In the *decoder* of the proposed DSTM scheme, with $\mathbf{U_t}$ in (4) set to $\mathbf{C_{Q4}}$ in (12), the decision metrics in (4) can be simplified to:

$$\{\hat{c}_1, \hat{c}_4\} = \underset{\{c_1, c_4\} \in \mathcal{M}}{\arg\max} \left[ \sum_{i=1,4} \mathrm{Re}\left\{\mathrm{tr}\left(\mathbf{R}_t^H \mathbf{R}_{t-1} \mathbf{A}_i\right)\right\} c_i^R + \sum_{i=1,4} \mathrm{Re}\left\{\mathrm{tr}\left(\mathbf{R}_t^H \mathbf{R}_{t-1} j\mathbf{B}_i\right)\right\} c_i^I \right]$$
$$\{\hat{c}_2, \hat{c}_3\} = \underset{\{c_2, c_3\} \in \mathcal{M}}{\arg\max} \left[ \sum_{i=2,3} \mathrm{Re}\left\{\mathrm{tr}\left(\mathbf{R}_t^H \mathbf{R}_{t-1} \mathbf{A}_i\right)\right\} c_i^R + \sum_{i=2,3} \mathrm{Re}\left\{\mathrm{tr}\left(\mathbf{R}_t^H \mathbf{R}_{t-1} j\mathbf{B}_i\right)\right\} c_i^I \right]$$
(16)

where $\mathbf{A}_k$ and $\mathbf{B}_k$ are the dispersion matrices of $\mathbf{C_{Q4}}$ [9].

As shown in (16), the proposed DSTM scheme can be decoded by the joint detection of two complex symbols ($c_1$ and $c_4$, or $c_2$ and $c_3$), and the two decision metrics can be computed separately. So the proposed unitary DSTM scheme is double-symbol decodable, just like its coherent counterpart in [8], and it has a lower decoding complexity than the DSTM schemes in [1- 6], which generally require a larger joint detection search space dimension. This will be elaborated in Section V, when we present performance results of our proposed codes.

*C. Design of Specific Joint Constellation Set*

In this section, we propose a pairwise constellation set $\mathcal{M}$ that satisfies all three requirements in (15) with good scalability in spectral efficiency. The proposed constellation set $\mathcal{M}$ is:

$$\{a_k, b_k\} = \begin{cases} a_k = \exp[j(2k\pi/M)]/\sqrt{2} \\ b_k = 0 \end{cases} \quad \text{for } 1 \leq k \leq L/2$$

$$\{a_k, b_k\} = \begin{cases} a_k = 0 \\ b_k = \exp[j(2(k-L/2)\pi/M + \theta)]/\sqrt{2} \end{cases} \quad \text{for } L/2 < k \leq L \quad (17)$$

where $M = L/2$ is an integer, and $\theta$ is a real number between 0 and $2\pi/M$.

Note that in (17), the parameter $M$ is related to the spectral efficiency *Eff* by *Eff* = $2(\log_2 2M)/N_T$ (since $L = 2M$) for a full-rate QO-STBC. Hence a unitary DSTM scheme with a wide range of spectral efficiency can be systematically designed from (17) by adjusting $M$. The parameter $\theta$ provides an extra degree of freedom to maximize the diversity and coding gain of the resultant unitary DSTM scheme.

*Theorem 1*: For the unitary DSTM constellation set defined in (17), the optimum value of $\theta$ (in the sense of the Performance Optimization Criterion (15)(iii)) is $\pi/M$ when $M$ is even, and is $\pi/2M$ or $3\pi/2M$ when $M$ is odd.

*Proof of Theorem 1*: The proof is given in four cases, considering different values of $k$ and $l$.

*Case 1: $1 \leq k, l \leq L/2$, and $k \neq l$.*

Since $b_k$ and $b_l$ are always zero in this case, $\Delta b_{kl}$ is always zero, but not $\Delta a_{kl}$ (since $k \neq l$). Hence the determinant value in (15)(iii) can never be zero, and its value is independent of $\theta$. This implies that, in this case, full diversity is always achieved, and the coding gain does not depend on $\theta$.

*Case 2: $L/2 < k, l \leq L$ and $k \neq l$.*

The proof is similar to Case 1 with the roles of $\Delta a_{kl}$ and $\Delta b_{kl}$ interchanged.

*Case 3:  $1 \leq k \leq L/2$ and $L/2 < l \leq L$.*

For this case, the determinant value in (15)(iii) can be simplified to:

$$\det = \frac{1}{16} \begin{bmatrix} (|\exp[j(2k\pi/M)] + \exp[j(2(l-L/2)\pi/M + \theta)]|^2) \times \\ (|\exp[j(2k\pi/M)] - \exp[j(2(l-L/2)\pi/M + \theta)]|^2) \end{bmatrix}^2 \quad (18)$$

$$= \left( \sin^2 \left( \frac{2\pi k}{M} - \frac{2\pi m}{M} - \theta \right) \right)^2 \quad 1 \leq k, m \triangleq l - \frac{L}{2} \leq M \text{ where } M \triangleq \frac{L}{2}$$

$$= \sin^4 \left( \frac{2(n-p)\pi}{M} \right) \quad 0 \leq n \triangleq k - m \leq M - 1$$

where $k, l, m, n$ are integers, and $p \triangleq M\theta/2\pi$ is a real number between 0 and 1 inclusively.

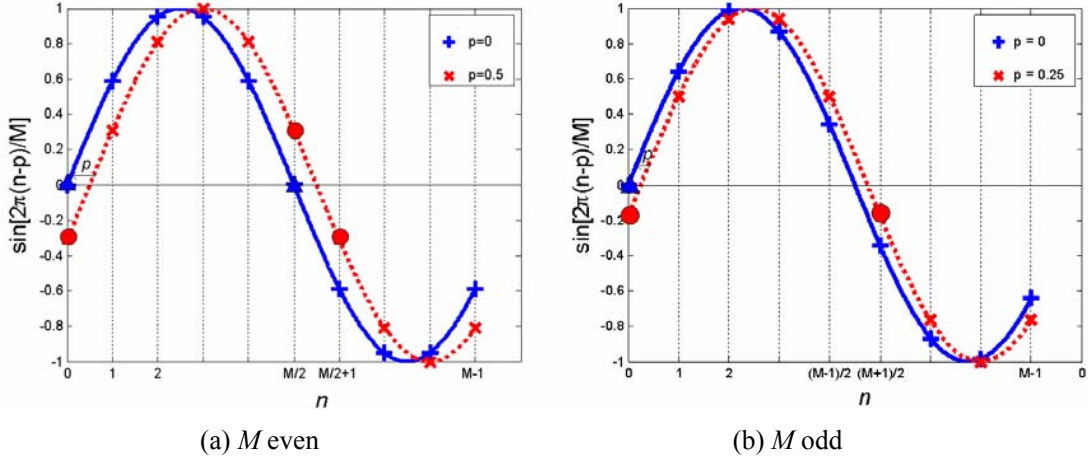

(a) $M$ even  (b) $M$ odd

Figure 1 Optimization of constellation rotation angle $\theta = 2\pi p/M$

To maximize the determinant value in (18), we first consider even values of $M$. As shown by the *triangular* markers in Figure 1(a), the function $\sin[2n\pi/M]$ (solid line) is zero if $n = 0$ or $M/2$. This results in a zero determinant value in (18) and the resultant unitary DSTM will not deliver full diversity. In order to achieve full diversity and maximum coding gain, a right shift $p$ can be introduced to obtain the function $\sin[2(n-p)\pi/M]$ (dashed line) such that its minimum *absolute* values at integer values of $n$ are non-zero and maximized, as shown by the *circular* markers in Figure 1(a). Clearly, this optimum point is reached when $p = 0.5$, which corresponds to $\theta = \pi/M$.

Similarly, as indicated by the circular markers in Figure 1(b), the optimum *p* value for odd values of *M* is 0.25 or 0.75, which corresponds to $\theta = \pi/2M$ or $3\pi/2M$.

*Case 4:  $1 \leq l \leq L/2$ and $L/2 < k \leq L$.*

The proof is similar to Case 3.

In summary, as the determinant value in (15)(iii) does not depend on $\theta$ for Cases 1 and 2, and the optimum $\theta$ value has been derived for Cases 3 and 4, *Theorem 1* is proved. ∎

There are a few important points to take note about the pairwise constellation set proposed in (14) and its optimum design stated in *Theorem 1*:

1. The constellation set specified in (14) is not the same as the conventional PSK with constellation rotation described in [9-13]. It is a special "pairwise/joint" constellation set that assigns 2 constellation points to 2 code symbols at a time, i.e. $c_1$ and $c_4$, or $c_2$ and $c_3$ in $\mathbf{C_{Q4}}$. In contrast, coherent QO-STBC typically assigns constellation points to individual codeword symbols independently (i.e. symbol by symbol), not 2 symbols at a time. Therefore our results in *Theorem 1* is fundamentally different from the constellation rotation results reported in [9-13], as *Theorem 1* pertains specifically to the proposed pairwise constellation set (14).

2. The "zero" symbols in $a_k$ or $b_k$ in (17) do not reduce the code rate of the proposed DSTM scheme by half. This is because every constellation pair $\{a_k, b_k\}$ in (17) represents 2 code symbols in the DSTM codeword, hence the "zero" symbols in $a_k$ or $b_k$ actually carry information – they are not null symbols. Since there are 4 code symbols in $\mathbf{C_{Q4}}$ and they will be represented by 2 pairs of complex constellation points drawn from (17), the proposed DSTM scheme effectively transmit 4 complex symbols in 4 symbol times, hence its code rate remains as 1, which is the same as the original $\mathbf{C_{Q4}}$.

## V. PERFORMANCE RESULTS

We now compare our two proposed double-symbol-decodable unitary DSTM schemes (one based on O-STBC with spherical code, the other based on QO-STBC with optimized joint constellation set specified in (17) and *Theorem 1*) against each other, as well as against existing unitary DSTM schemes. In Table 2, we compare the coding gain and decoding complexity of our proposed DSTM schemes against those based on square O-STBC [7] and group codes [1, 2] for four transmit antennas. We can see that both proposed unitary DSTM schemes provide higher coding gain than the O-STBC and group-code DSTM schemes, at both spectral efficiency values of 1.5bps/Hz and 2bps/Hz. Our proposed DSTMs also have lower decoding complexity than the group-code DSTM, because they can be decoded with two parallel decoders, each with a decoding search space dimension of 8 at 1.5bps/Hz and 16 at 2 bps/Hz. Although the unitary DSTM based on rate-3/4 square O-STBC with 64PSK [7] has an even lower decoding complexity at a spectral efficiency 1.5bps/Hz, it suffers from a much lower coding gain. At a spectral efficiency of 2 bps/Hz, our proposed DSTMs have higher coding gains than, and equal decoding search space dimension as, the DSTM based on rate-1/2 square O-STBC with 16PSK [7]. It should also be noted that the use of joint modulation enables our DSTM schemes to flexibly support various spectral efficiency values. This is not possible for the O-STBC and QO-STBC DSTM schemes proposed in [15,16].

Table 2 Comparison of coding gain and decoding search space per decoder
of unitary DSTM schemes for four transmit antennas

| Spectral efficiency | Unitary DSTM scheme | Constellation | Coding gain | Number of parallel decoders | Decoding search space per decoder |
|---|---|---|---|---|---|
| 1.5 | [1, 2] | 64PSK | 1.85 | 1 | 64 |
| **1.5** | **Proposed scheme with rate-3/4 O-STBC (Section III)** | **Spherical code (3d, 8 points)** | **2.95** | **2** | **8** |
| **1.5** | **Proposed scheme with rate-1 QO-STBC (Section IV)** | **(17) with $M = 4$, $\theta = \pi/4$** | **2.83** | **2** | **8** |
| 1.5 | Rate-3/4 O-STBC [7] | QPSK | 2.70 | 3 | 4 |
| 2 | [1, 2] | 256PSK | 0.78 | 1 | 256 |
| **2** | **Proposed scheme with rate-3/4 O-STBC (Section III)** | **Spherical code (3d, 16 points)** | **1.55** | **2** | **16** |
| **2** | **Proposed scheme with rate-1 QO-STBC (Section IV)** | **(17) with $M = 8$, $\theta = \pi/8$** | **1.17** | **2** | **16** |

| 2 | Rate-1/2 O-STBC [7] | 16-PSK | 0.31 | 2 | 16 |

In Figure 2, we compare the block error rate (BLER) performance of our proposed DSTM schemes with those reported in [6, 7] for four transmit and one receive antennas. Compared with the 2 bps/Hz DSTM based on rate-1/2 square O-STBC [7], both our proposed DSTM schemes have much better BLER performance, which agrees with the superior coding gain observation already made in Table 2. Compared with the Sp(2) DSTM scheme of spectral efficiency 1.94 bps/Hz and decoding search space dimension of 225 (obtained from [6] with $M=5$, $N=3$), our proposed DSTM schemes perform no more than 1dB worse, but both have much smaller decoding search space dimension of 16 and a slightly higher spectral efficiency.

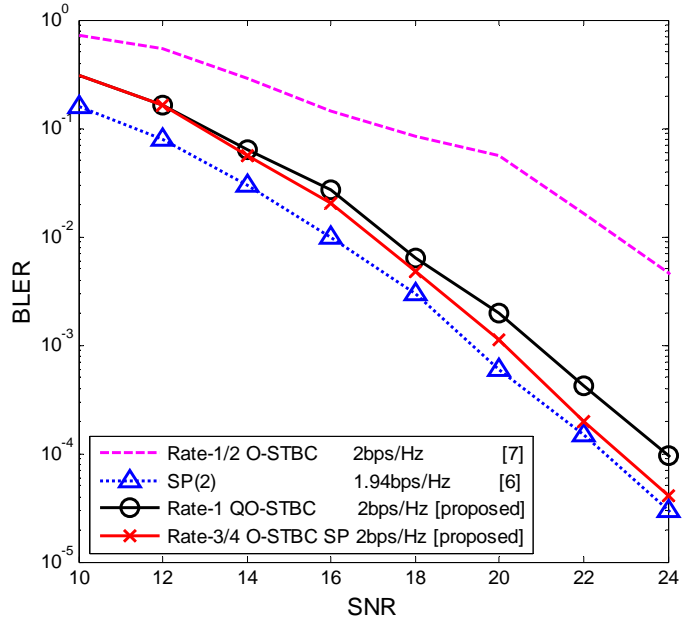

Figure 2 Block error rates of different DSTM schemes for four tx and one rx antennas

Table 3 compares the coding gain and decoding search space per decoder of three unitary DSTM schemes for eight transmit antennas. The first DSTM is our proposed DSTM based on a rate-1/2 O-STBC with spherical code. In this setting, we jointly modulate two complex symbols. In order to achieve a spectral efficiency of 1.5bps/Hz, we employ a spherical code of four dimensions (two complex symbols is equivalent to four real dimensions) and 64 points. It can be decoded by two parallel decoders, each with a decoding search space of 64. The second DSTM under comparison is our proposed DSTM based on rate-3/4 QO-STBC with the joint constellation set specified in (17) and *Theorem 1*. In this setting, we jointly modulate two complex symbols. This DSTM can be decoded by three parallel decoders, each with a search space of 16. The third DSTM under comparison is the DSTM based on rate-1/2 O-STBC from [7] that employs 8-PSK. Decoding it requires four parallel decoders, each with a search space of 8. Table 3 shows that both our proposed DSTM schemes have higher coding gain than the third scheme. Interestingly, among our two proposed schemes, the one based on O-STBC with spherical code has a higher coding gain, but also larger decoding search space, than that based on QO-STBC (unlike the case of four transmit antennas). In Table 3 we also demonstrate that our proposed QO-STBC DSTM scheme can be extended to eight transmit antennas and still maintains pair-wise decoding complexity. Such extension to eight transmit antennas is not possible for the non-linear DSTM reported in [16].

Table 3 Comparison of coding gains and decoding search space per decoder of unitary DSTM for eight transmit antennas

| Spectral efficiency | Unitary DSTM scheme | Constellation | Coding gain | Number of parallel decoder | Decoding search space per decoder |
|---|---|---|---|---|---|
| **1.5** | **Proposed scheme with rate-1/2 O-STBC (Section III)** | **Spherical code (4d, 64 points)** | **2.08** | **2** | **64** |
| **1.5** | **Proposed scheme with rate-3/4 QO-STBC (Section IV)** | (17) with $M = 8$, $\theta = \pi/8$ | **1.56** | **3** | **16** |
| 1.5 | Rate-1/2 O-STBC [7] | 8-PSK | 1.17 | 4 | 8 |

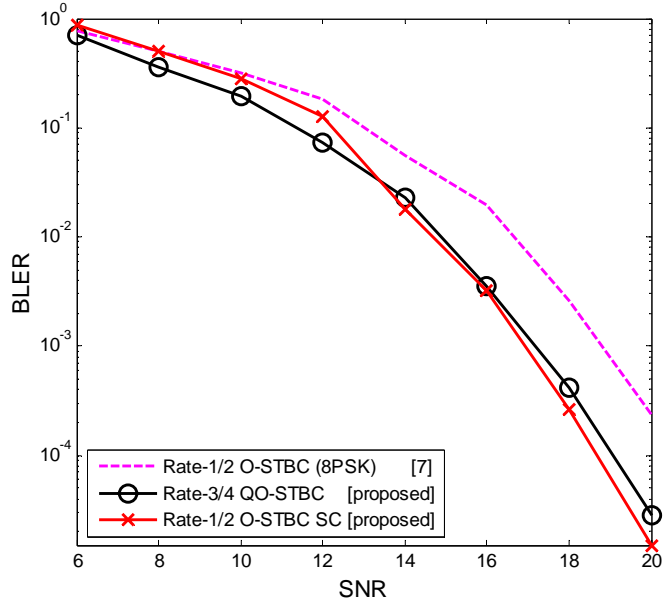

Figure 3 Block error rates of for eight tx and one rx antennas at 1.5 bits/sec/Hz

The decoding performance of the three DSTM schemes from Table 3 are compared in Figure 3. We can see that the best-performing scheme is our proposed DSTM based on rate-1/2 O-STBC with spherical code, followed by our proposed DSTM based on rate-3/4 QO-STBC, and lastly the DSTM based on rate-1/2 square O-STBC from [7]. This generally agrees with the coding gain ranking in Table 3, but we can observe that the BLER performance difference between our proposed O-STBC and QO-STBC schemes is not as large as what their coding gain difference may suggest.

## VI. CONCLUSIONS

Two new unitary differential space-time modulation (DSTM) schemes with low decoding complexity are proposed. The first design is based on O-STBC with joint constellation constructed from spherical code. The main idea is to jointly modulate multiple symbols using a set of joint constellation points constructed based on spherical codes The second design is based on unitary matrices constructed from double-symbol-decodable QO-STBC. The main idea is to force the QO-STBC codeword to be a unitary matrix by using pair-wise symbol modulation with a specially designed constellation set. Our proposed unitary DSTM schemes have much smaller decoding search space per decoder than the DSTM schemes reported in $[1-6]$, with comparable or better coding gain or decoding performance. Compared with the O-STBC DSTM reported in [7], at a block error rate of $10^{-3}$ or lower, both our proposed DSTM schemes give more than 3dB and 1dB decoding performance gain for 4 and 8 transmit antennas respectively, with only a slight increase in decoding complexity.

## ACKNOWLEDGMENT

The authors would like to thank the anonymous reviewers and the associate editor for their comments that greatly improve this paper.

## REFERENCES

[1] B. L. Hughes, "Differential space-time modulation", *IEEE Trans. on Info. Theory*, vol: 46, Nov. 2000, pp. 2567–2578.

[2] B. L. Hughes, "Optimal space-time constellations from groups", *IEEE Trans. on Info. Theory*, vol: 49, Feb. 2003, pp. 401– 410.

[3] B. M. Hochwald and W. Sweldens, "Differential unitary space-time modulation", *IEEE Trans. on Comms.*, vol: 48, Dec. 2000, pp. 2041–2052.

[4] V. Tarokh and H. Jafarkhani, "A differential detection scheme for transmit diversity", *IEEE Journal on Selected Areas in Comms.*, vol: 18, July 2000, pp. 1169–1174.

[5] B. Hassibi and B. M. Hochwald, "Cayley differential unitary space-time codes", *IEEE Trans. on Info. Theory*, 2002, pp. 1485 – 1503.

[6] Y. Jing and B. Hassibi, "Design of fully-diverse multi-antenna codes based on Sp(2)", *ICASSP 2003*, pp. 33-36.

[7] G. Ganesan and P. Stoica, "Differential modulation using space-time block codes", *IEEE Signal Processing Letters*, vol: 9, Feb 2002, pp. 57 –60.

[8] H. Jafarkhani, "A quasi-orthogonal space-time block code", *IEEE Trans. on Comms.,* vol: 49, Jan. 2001, pp. 1-4.

[9] C. Yuen, Y. L. Guan, and T. T. Tjhung, "Full-Rate Full-Diversity STBC with Constellation Rotation", *VTC-Spring 2003,* pp. 296 –300.

[10] O. Tirkkonen, "Optimizing STBC by Constellation Rotations", *FWCW 2001*, pp. 59-60.

[11] N. Sharma and C. B. Papadias, "Improved quasi-orthogonal codes through constellation rotation", *IEEE Trans. on Comms.*, vol: 51, March 2003, pp. 332- 335.

[12] W. Su and X. Xia, "Signal constellations for quasi-orthogonal space-time block codes with full diversity", *IEEE Trans. on Information Theory*, vol. 50, pp. 2331 – 2347, Oct. 2004.

[13] D. Wang and X. -G. Xia, "Optimal Diversity Product Rotations for Quasi-Orthogonal STBC with MPSK Symbols", *IEEE Communications Letters*, vol. 9, pp. 420 – 422, May 2005.

[14] B. Hassibi and B. M. Hochwald, "High-Rate Codes that are Linear in Space and Time", *IEEE Trans. on Info. Theory*, vol: 48, Jul 2002, pp. 1804 –1824.

[15] Y. Zhu and H. Jafarkhani, "Differential modulation based on quasi-orthogonal codes", *WCNC 2004*.

[16] A. R. Calderbank, S. Diggavi, S. Das, and N. Al-Dhahir, "Construction and analysis of a new 4x4 orthogonal space-time block code", *IEEE ISIT 2004*, pp. 309.

[17] G. Ganesan and P. Stoica, "Space-time block codes: a maximum SNR approach", *IEEE Trans. on Information Theory*, vol. 47, pp.1650-1656, May 2001.

[18] N. J. A. Sloane, with the collaboration of R. H. Hardin, W. D. Smith and others, "Tables of Spherical Codes", published electronically at www.research.att.com/~njas/packings/

**Appendix A**

**Optimal 16-Point 3-Dimension Spherical Code**

|    | $a_k$        | $b_k$        | $c_k$         |
|----|--------------|--------------|---------------|
| 1  |  0.089527456 |  0.681333248 | -0.166642852; |
| 2  | -0.418469889 |  0.545080831 |  0.166642852; |
| 3  |  0.057360813 |  0.436534568 |  0.5533058;   |
| 4  | -0.268116333 |  0.349236773 | -0.5533058;   |
| 5  | -0.681333248 |  0.089527456 | -0.166642852; |
| 6  | -0.545080831 | -0.418469889 |  0.166642852; |
| 7  | -0.436534568 |  0.057360813 |  0.5533058;   |
| 8  | -0.349236773 | -0.268116333 | -0.5533058;   |
| 9  | -0.089527456 | -0.681333248 | -0.166642852; |
| 10 |  0.418469889 | -0.545080831 |  0.166642852; |
| 11 | -0.057360813 | -0.436534568 |  0.5533058;   |
| 12 |  0.268116333 | -0.349236773 | -0.5533058;   |
| 13 |  0.681333248 | -0.089527456 | -0.166642852; |
| 14 |  0.545080831 |  0.418469889 |  0.166642852; |
| 15 |  0.436534568 | -0.057360813 |  0.5533058;   |
| 16 |  0.349236773 |  0.268116333 | -0.5533058.   |